\documentclass[prd,nofootinbib ,preprintnumbers, superscriptaddress, 14pt]{revtex4-2}

\usepackage{lineno}

\usepackage{amsmath,mathrsfs}
\usepackage[maxfloats=100]{morefloats}
\usepackage{color,xcolor}
\usepackage[normalem]{ulem}
\usepackage{subfig}
\usepackage{amsfonts,amssymb,epsfig}
\usepackage{bbold}
\usepackage{bm}
\usepackage{graphicx}
\usepackage{dcolumn}
\usepackage{multirow}
\usepackage{tabularx}
\usepackage{mathtools}
\usepackage{slashed}
\usepackage{booktabs}
\usepackage{float}
\usepackage[colorlinks=true,
            linkcolor=blue,
            urlcolor=blue,
            citecolor=red,          
            bookmarks=true,
            bookmarksnumbered=true,
            breaklinks=true,
            pdfpagemode=Fullscreen,
            pdfstartview=FitBH]{hyperref}

\usepackage{tikz}
\usepackage[compat=1.1.0]{tikz-feynhand}
\usetikzlibrary{patterns, math}
\usetikzlibrary{positioning}

\makeatletter
\renewcommand\normalsize{%
    \@setfontsize\normalsize{12pt}{14pt} 
}

\newcommand{\PZ}{\ensuremath{\mathrm{Z}}}
\newcommand{\PH}{\ensuremath{\mathrm{H}}}

\newcommand{\MGMCatNLO}{MadGraph5\_aMC@NLO} 

\newcommand{\delphes}{{\sc Delphes}}

\newcommand{\met}{\ensuremath{\slashed{E}_T}}

\newcommand{\PTeV}{\ensuremath{\mathrm{TeV}}}
\newcommand{\PGeV}{\ensuremath{\mathrm{GeV}}}

\newcommand{\mc}{\mathcal}
\def\Tr{\mbox{Tr}\,}
\newcommand{\madgraph}{\textsc{MadGraph5\_aMC@NLO}}
\newcommand{\ket}[1]{\left| #1 \right\rangle}
\newcommand{\bra}[1]{\left\langle #1 \right|}

\begin{document}

\title{Testing Bell inequalities and probing quantum entanglement at CEPC}

\author{Youpeng \surname{Wu}}
\email[]{youpeng@stu.pku.edu.cn}
\affiliation{State Key Laboratory of Nuclear Physics and Technology, School of Physics, Peking University, Beijing, 100871, China}

\author{Ruobong \surname{Jiang}}
\email[]{ruobing@stu.pku.edu.cn}
\affiliation{State Key Laboratory of Nuclear Physics and Technology, School of Physics, Peking University, Beijing, 100871, China}

\author{Alim \surname{Ruzi}}
\email[]{alim.ruzi@pku.edu.cn}
\affiliation{State Key Laboratory of Nuclear Physics and Technology, School of Physics, Peking University, Beijing, 100871, China}

\author{Yong \surname{Ban}}
\email[]{bany@pku.edu.cn}
\affiliation{State Key Laboratory of Nuclear Physics and Technology, School of Physics, Peking University, Beijing, 100871, China}

\author{Qiang \surname{Li}}
\email[]{qliphy0@pku.edu.cn}
\affiliation{State Key Laboratory of Nuclear Physics and Technology, School of Physics, Peking University, Beijing, 100871, China}

\begin{abstract}
    We study quantum entanglement and test violation of Bell-type inequality at the Circular Electron Positron Collider (CEPC), which is one of the most attractive future collides. It's a promising particle collider designed to search new physics, make Standard Model (SM) precision measurements, and serving as a Higgs factory.  Our study is based on a fast simulation of the $\PZ$ boson pair production from Higgs boson decay at  $\sqrt{s} = 250$ \PGeV. The detector effects are also included in the simulation. The spin density matrix of the joint $\PZ\PZ$ system is parametrized using irreducible tensor operators and reconstructed from the spherical coordinates of the decay leptons. To test Bell inequalities, we construct observable quantities for the \(H \rightarrow ZZ*\) process in CEPC by using the (Collins-Gisin-Linden-Massar-Popescu) CGLMP inequality, whose value is determined from the density matrix of the Z boson pairs. The sensitivity of the Bell inequality violation is observed with more than 1$\sigma$ and the presence of the quantum entanglement is probed with more than 2$\sigma$ confidence level.
\end{abstract}

\maketitle
\section{Introduction}
\label{sec:intro}

A.~Einstein, B.~Podolsky and N.~Rosen,
Since the advent of quantum mechanics, observation of quantum entanglement~\cite{Horodecki:2009zz} in a paired quantum-mechanical system has been a continuous and ongoing research topic. The concept of entanglement was first introduced by Schr\"odinger in 1935, and the famous EPR paradox was proposed by Einstein, Podolsky, and Rosen in 1935~\cite{Einstein:1935rr}. The EPR paradox is a thought experiment that challenges the completeness of quantum mechanics. In 1964, Bell proposed a set of inequalities~\cite{Bell:1964kc} to test a local hidden variable  theory, which is a class of theories that can reproduce the predictions of quantum mechanics. Violation of Bell inequalities implies that the local hidden variable theory is not valid, and quantum mechanics is a complete theory. In 1978, Clauser \rm{et al.} proposed a more practical form of Bell inequality for two-qubit system, which is called CHSH inequality~\cite{Clauser:1978ng}. This inequality has been tested for decades in experiments designed for quantum entanglement study. The CGLMP inequalities, on the other hand, are a generalization of the CHSH inequalities for two-qutrit system and can be used to test the local hidden variable theory in a more efficient way. Quantum entanglement has been successfully observed in two-outcome measurements with correlated photon pairs~\cite{aspect1982experimental,Freedman:1972zza}

The study of the quantum entanglement and testing Bell inequality violation on the colliders is gaining a wide range of interest in high energy physics community. Recently, a number of proposals have been made to test Bell inequalities and probe quantum entanglement through quantum state tomography of top-quark pairs~\cite{Fabbrichesi:2021npl,Afik:2020onf, Severi:2021cnj, Aguilar-Saavedra:2022uye, Afik:2022dgh,Afik:2022kwm, Han:2023fci, Dong:2023xiw} as well as heavy lepton pair production at a lepton collider and other future collider~\cite{Ehataht:2023zzt,Ma:2023yvd,Gray:2021jij}. Large Hadron Collider (LHC) experiments  have also been conducted to test Bell inequalities through measuring the spin polarization of top and anti-top quarks~\cite{CMS:2019nrx}, and the entanglement between top and anti-top quark  events has been observed~\cite{CMS:2024pts,ATLAS:2023fsd}, also in LCHb and Belle II experiments observed violation of the Bell inequality in B meson decay~\cite{Fabbrichesi:2024rec}. Endeavors to probe quantum entanglement have been made not only at the LHC through massive gauge boson pairs~\cite{Morales:2023gow,Morales:2024jhj,Aguilar-Saavedra:2024whi,Grossi:2024jae} but also at future colliders, such as a Muon collider through the $\PH \rightarrow \PZ\PZ$ at $\PTeV$ collision energy. Analysis from a simulated Muon collider has shown that probing quantum entanglement and testing Bell inequality in pair of massive gauge boson production is also achievable~\cite{Ruzi:2024cbt}. Interestingly, quantum entanglement studies in the context of quantum field theories, such as Quantum Electro Dynamics (QED), have shown quantum entanglement might even take place in two-two scattering process~\cite{Sinha:2022crx,Fedida:2022izl,Thaler:2024anb}

The CEPC is a proposed circular electron-positron collider in China, which is designed to study the Higgs boson and other particles with high precision. The main processes that generate the Higgs boson at the CEPC are following three processes: Higgsstrahlung(\(e^+e^-\rightarrow \PZ\PH\)), WW fusion(\(e^+e^-\rightarrow\nu\bar{\nu}\PH\)), and ZZ fusion(\(e^+e^-\rightarrow e^+ e^- \PH\))~\cite{Kiuchi:2021roe,Chen:2016zpw}. And Higgsstrahlung is the main dominant process at the center-of-mass energy of 250 $\PGeV$, which is the signal process we choose in this paper.
The CEPC is an ideal platform to test Bell inequalities and probe quantum entanglement, on the one hand, leptonic collision provides very clean backgrounds and simple final states to test the Bell inequalities, on the other hand, the Higgs boson decay process \(\PH \rightarrow \PZ\PZ^{*}\) provides the conservation of angular momentum and spin polarization, so the Z bosons in final state are entangled. 

\begin{figure}[htb]
    \centering
    (a)
\begin{tikzpicture}
    \begin{feynhand}
        \vertex (e+) at     (-1,2) {\(e^+\)};
        \vertex (e-) at     (-1,-2) {\(e^-\)};
        \vertex (eez) at    (0,0);
        \vertex (zzh) at    (1,0);
        \vertex (zdec) at   (2,-1);
        \vertex (hzz) at    (2,1);
        \vertex (z1) at     (3.2,1.5);
        \vertex (z2) at     (3.2,0.5);
        \vertex (q1) at     (3.2,-1.5) {\(j\)};
        \vertex (q2) at     (3.2,-0.5) {\(j\)};
        \vertex (l1) at     (4.5,1.8)   {\(l_1\)};
        \vertex (l2) at     (4.5,1.2) {\(l_2\)};
        \vertex (l3) at     (4.5,0.8) {\(l_3\)};
        \vertex (l4) at     (4.5,0.2)   {\(l_4\)};
        \propag[fer] (e-) to (eez);
        \propag[antfer] (e+) to (eez);
        \propag[bos] (eez) to [edge label=\(Z\)](zzh);
        \propag[sca] (zzh) to [edge label=\(H\)](hzz);
        \propag[bos] (zzh) to [edge label=\(Z\)](zdec);
        \propag[fer] (zdec) to (q1);
        \propag[antfer] (zdec) to (q2);
        \propag[bos] (hzz) to [edge label=\(Z_1\)](z1);
        \propag[bos] (hzz) to [edge label=\(Z_2\)](z2);
        \propag[fer] (z1) to (l1);
        \propag[antfer] (z1) to (l2);
        \propag[fer] (z2) to (l3);
        \propag[antfer] (z2) to (l4);
    \end{feynhand}
\end{tikzpicture}
    (b)
\begin{tikzpicture}
    \begin{feynhand}
        \vertex (e+) at     (-1,2) {\(e^+\)};
        \vertex (e-) at     (-1,-2) {\(e^-\)};
        \vertex (eez) at    (0,0);
        \vertex (zzh) at    (1,0);
        \vertex (zdec) at   (2,-1);
        \vertex (hzz) at    (2,1);
        \vertex (z1) at     (3.2,1.5);
        \vertex (z2) at     (3.2,0.5);
        \vertex (q1) at     (3.2,-1.5) {\(\nu\)};
        \vertex (q2) at     (3.2,-0.5) {\(\bar{\nu}\)};
        \vertex (l1) at     (4.5,1.8)   {\(l_1\)};
        \vertex (l2) at     (4.5,1.2) {\(l_2\)};
        \vertex (l3) at     (4.5,0.8) {\(l_3\)};
        \vertex (l4) at     (4.5,0.2)   {\(l_4\)};
        \propag[fer] (e-) to (eez);
        \propag[antfer] (e+) to (eez);
        \propag[bos] (eez) to [edge label=\(Z\)](zzh);
        \propag[sca] (zzh) to [edge label=\(H\)](hzz);
        \propag[bos] (zzh) to [edge label=\(Z\)](zdec);
        \propag[fer] (zdec) to (q1);
        \propag[antfer] (zdec) to (q2);
        \propag[bos] (hzz) to [edge label=\(Z_1\)](z1);
        \propag[bos] (hzz) to [edge label=\(Z_2\)](z2);
        \propag[fer] (z1) to (l1);
        \propag[antfer] (z1) to (l2);
        \propag[fer] (z2) to (l3);
        \propag[antfer] (z2) to (l4);
    \end{feynhand}
\end{tikzpicture}
    \caption{Signal processes we choose to test Bell inequalities for CEPC. (a) semi-leptonic channel \(e^+e^- \rightarrow \PZ\PH, \PZ\rightarrow jj\), (b) pure-leptonic channel \(e^+e^- \rightarrow \PZ\PH, \PZ\rightarrow\nu\bar{\nu}\).}
    \label{fig:signal}
\end{figure}
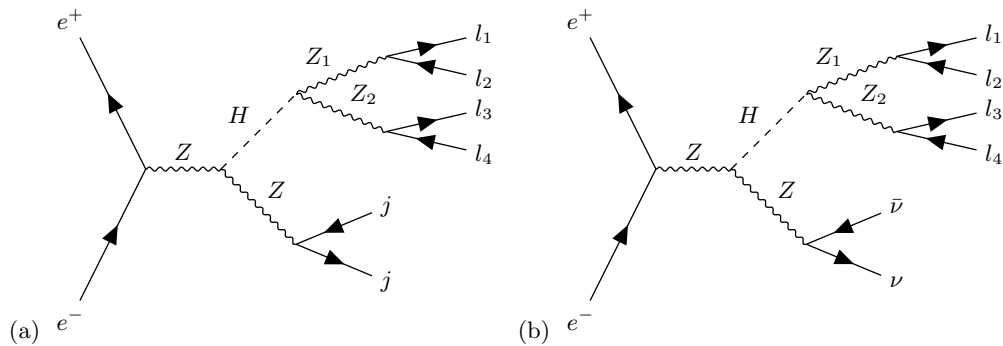

we propose a method to test Bell inequalities at the CEPC using the Higgs boson decay process \(H \rightarrow ZZ^*\). Our signal is $e^{+} e^{-} \rightarrow \PZ \PH, (\PH \rightarrow \ell^{+} \ell^{-} \PZ, \PZ \rightarrow \ell^{+} \ell^{-})$, we also divide the process into pure-leptonic channel and semi-leptonic channle depending on the Z boson's decay. We construct an observable $\mc{I}_{3}$ for the \(\PH \rightarrow \PZ\PZ^{*}\) process that can be used to test the CGLMP inequality. We apply identical coordinate system to measure the spin polarization observable as in references~\cite{Bernreuther_2015yna,Aguilar-Saavedra_2022wam}. First of all, a set of coordinates are set up in Higgs boson rest frame from which another set of coordinate in $Z$ boson rest frame is obtained via a rotation. Then a Lorentz boost is applied on the momentum of final leptons so that they are now in the $\PZ$ boson rest frame. Finally, the angular coordinates of the leptons, $(\theta_1^-, \phi_1^-)$ for negative-charged lepton from $\PZ_1$ decay and $(\theta_2^-, \phi_2^-)$ for negative-charged lepton from $\PZ_2$ decay can be obtained and used to determine coefficients of the density matrix of the joint $ZZ$ system. We use simulate the signal process using publicly available Monte-Carlo program \madgraph.~The observables namely \(\mc{I}_3\) of the CGLMP inequality and coefficients that quantify quantum enatangelment are obtained from those simulated data. We also study the possible backgrounds for this process with a collision energy at 250 \PGeV, and $\mathcal{L}=50ab^{-1}$  .

\section{Theoretical approach and observable construction}
\label{sec:theory}
\subsection{Bell Inequalities}

For a two-qutrit system of $\PZ$, we extend the CGLMP inequality, considering two observers, A and B, each capable of measuring the polarization of a $\PZ$ boson along three distinct directions. This setup resembles a typical Bell-type experiment, in which we statistically analyze the measurement outcomes from both observers to obtain the probability distribution for these outcomes. With a simulated experiment, we use a density matrix to represent the mixed state of the $\PZ$ boson pair. Leveraging the properties of the density matrix, we can construct an observable to test the Bell inequalities, with the expectation value given by $\Tr[\rho \mc B]$ where $\mc B$ denotes the Bell operator.

The general form of Bell inequality is deduced from a local-hidden-variable theory. Basically, it means we can write the individual probabilities of an outcome of a measurements as \(P(A_1B_1|AB,\lambda)= P(A_1|A,\lambda)P(B_1|B,\lambda)\). Without loss of generality, Bell inequality for two-qudit system with total dimension d is given as 
\begin{align}
    \nonumber
    \mc{I}_d=\sum_{k-0}^{[d/2]-1}(1-\frac{2k}{d-1})&\{+[P(A_1=B_1+k)+P(B_1=A_2+k+1)+P(A_2=B_2+k)\\
    \nonumber
    & +P(B_2=A_1+k)-\left[ P\left( A_1=B_1-k-1 \right) +P\left( B_1=A_2-k \right)\right.\\
    &  +P\left( A_2=B_2-k-1 \right) \left.+ P(B_2=A_1-k-1)\right]\}
\end{align}
It can provide a more robust way to test the local hidden variable theory. In our case, cause the Z boson is a spin-1 particle, we can use the 3 dimensional form of the CGLMP inequality, given by
\begin{align}
    \nonumber
    \mc{I}_3 = &P(A_1 = B_1) + P(B_1 = A_2+1) + P(A_2 = B_2) + P(B_2 = A_1)\\
    &-[P(A_1 = B_1 - 1) + P(B_1 = A_2) + P(A_2 = B_2 - 1) + P(B_2 =A_1 - 1)].
    \label{eq:i3}
\end{align}
For classical local hidden variable theory, \(\mc{I}_3 \leq 2\) can be derived ~\cite{Collins:2002sun}. It is generally considered violation of Bell inequality is a very strong evidence of quantum entanglement,  implying that the local hidden variable theory is not valid and quantum mechanics is a complete theory. 

\subsection{Density Matrix of $H\rightarrow ZZ^*$}

We construct the density matrix of the Z boson pairs from the di-boson Higgs decay. The Z boson is a spin-1 particle,  for two Z bosons, the density matrix \(\rho\) on the 9 dimensional Hilbert space. The basis vectors we choose is the eigenstates of the momentum operator \(J_z\) by defining the z-axis along the momentum of the Z boson. The $Z$ boson pairs are produced from Higgs decay, and the spin component is conserved in the momentum direction in the CM frame. Therefore, the ZZ state can only lie in one of 3 joint states, i.e., \({\ket{l_1l_2}\in\left\{\ket{-+}, \ket{00}, \ket{+-}\right\}}\), where \(l_1\) and \(l_2\) are the spin states of two Z bosons. By definition, the density matrix of a two-qutrit ZZ system is written in the tensor-product form
\begin{align}
    \rho=\sum p_{l_1l_2}\ket{l_1l_2}\bra{l_1l_2},
\end{align} 
where $p_{l_1l_2}\geq 0$ and $\sum p_{l_1l_2}=1$. 

For more general case, we determine z-axis as the direction of on-shell Z boson in Higgs center-of-mass frame, and the XOY plane contains unit vector \(\hat{x}\) in laboratory frame. Of course, the x-axis unit vector is vertical to the z-axis so that we define as \(\hat{r}=\mathrm{sign}(\cos \Theta)(\hat{x}-\hat{k}\cos \Theta)/\sin\theta\), where \(\Theta\) is the angle between the z-axis (unit vector of the momentum of the on-shell Z boson) and the beam direction. We can naturally obtain the present frame of the coordinate system  by means of the right-hand spiral theorem, and this convention will be employed for all subsequent content.

The form of the density matrix describing the polarization state of the
two-qutrit system formed by two spin-1 bosons can generally be parameterized using $3\times 3$
matrices composed of either Gell-Mann matrices~\cite{Fabbrichesi:2023cev,Barr:2024djo} or spin-1 operators. Using Gell-Mann matrices to represent the density matrix is one of the possible parameterization. There is another simple yet effective way to parameterize the density matrix composed of linear combination of irreducible tensor operators ~\cite{Rahaman:2021fcz,Aguilar-Saavedra:2022wam}
\begin{align}
    \rho =\frac{1}{9}\left[ \mathbb{1} _3\otimes \mathbb{1} _3+A_{LM}^{1}T_{M}^{L}\otimes \mathbb{1} _3+A_{LM}^{2}\mathbb{1} _3\otimes T_{M}^{L}+C_{L_1\mathrm{M}_2\mathrm{L}_2\mathrm{M}_2}T_{M_1}^{L_1}\otimes T_{M_2}^{L_2} \right] 
    \label{eq:rho}
\end{align}
where \(T_M^L\) are the irreducible tensor operators complying with \(\Tr[T_M^L(T_M^L)^{\dagger}]=3\). These operators are summed over the indices \(M=-L,-L+1,\cdots,L\) and \(L=1,2\). The coefficients this fomular \(A_{LM},C_{L_1M_1L_2M_2}\) are the correlation coefficients which can be calculated from the density matrix. For the most general case, the density matrix should be a \(9\times9\). However, the Z boson pairs are produced from the Higgs boson decay, considering the decay density matrix of a Z boson into charged leptons, the elements of the density matrix can be constrained form this condition.

So the key point is to extract the correlation coefficients \(A_{LM},C_{L_1M_1L_2M_2}\) from simulated or actual experimental data. For no we can obtain these parameters from the differential cross section of the signal process as our experimental data from CEPC. 
The differential cross section of the signal process \(ZZ \rightarrow l_1^+l_1^-l_2^+l_2^-\) can be written as~\cite{Rahaman:2021fcz}
\begin{align}
    \frac{1}{\sigma}\frac{d\sigma}{d\Omega_+d\Omega_-}=(\frac{3}{4\pi})^2\Tr [\rho(\Gamma_1\otimes\Gamma_2)],
\end{align} 
where \(\Gamma_1\) and \(\Gamma_2\) are the decay density matrix of the Z boson into charged leptons. \(\Omega\) is the solid angle given by the spherical coordinates of the final state leptons. The trace can be simplified further using the normalization property of the irreducible tensors  and making use of spherical harmonic functions \(Y_L^M(\theta,\phi)\). The differential cross section can be written as
\begin{align}
\nonumber
    \frac{1}{\sigma}\frac{d\sigma}{d\Omega_1d\Omega_2} =&\frac{1}{(4\pi)^2} [ 1+A_{LM}^1Y_L^M(\theta_1,\phi_1)+A_{LM}^2B_LY_L^M(\theta_2,\phi_2)\\
    & +C_{L_1M_1L_2M_2}B_{L_1}B_{L_2}Y_{L_1}^{M_1}(\theta_1,\phi_1)Y_{L_2}^{M_2}(\theta_2,\phi_2)
    ] ,
\end{align}
we use irreducible tensor operator's orthogonality to simplify the expression, 
where \(B_L\) is the constant \(B_1=-\sqrt{2\pi\eta_l},B_2=\sqrt{2\pi/5}\). Now we can use the orthogonality of spherical harmonic functions to simplify the expression. Integrated over the solid angle, we can get the correlation coefficients \(A_{LM},C_{L_1M_1L_2M_2}\) from the above equation:
\begin{align}
    \int \frac{1}{\sigma}\frac{d\sigma}{d\Omega_1d\Omega_2}Y_L^M(\Omega_j)d\Omega_j&=\frac{B_L}{4\pi}A_{LM}^j, \qquad j=1,2;\\
    \int \frac{1}{\sigma}\frac{d\sigma}{d\Omega_1d\Omega_2}Y_{L_1}^{M_1}(\Omega_1)Y_{L_2}^{M_2}(\Omega_1)d\Omega_1d\Omega_2 &= \frac{B_{L_1}B_{L_2}}{4\pi}C_{L_1M_1L_2M_2}.
    \label{eq:ACLM}
\end{align}
It is worth noting that the ZZ system is in the singlet state because the third component of the spin along the  boson momentum direction is conserved. This imposes strong constraint on the configuration of density matrix, including only nine  non-zero elements with the relation
\begin{equation}
    C_{2,2,2,-2} = \frac{1}{\sqrt{2}}A_{2, 0}^1.
\end{equation}

\subsection{Observable}

Our observable is constructed from CGLMP inequalities. We follow the formulation of the Bell operator for two-qutrit system in~\cite{Aguilar-Saavedra:2022wam}:
\begin{align}
    \nonumber
	\mathcal{B} &=\left[ \frac{2}{3\sqrt{3}}\left( T_{1}^{1}\otimes T_{1}^{1}-T_{0}^{1}\otimes T_{0}^{1}+T_{1}^{1}\otimes T_{-1}^{1} \right) +\frac{1}{12}\left( T_{2}^{2}\otimes T_{2}^{2}+T_{2}^{2}\otimes T_{-2}^{2} \right) \right.\\
	&\left. +\frac{1}{2\sqrt{6}}\left( T_{2}^{2}\otimes T_{0}^{2}+T_{0}^{2}\otimes T_{2}^{2} \right) -\frac{1}{3}(T_{1}^{2}\otimes T_{1}^{2}+T_{1}^{2}\otimes T_{-1}^{2})+\frac{1}{4}T_{0}^{2}\otimes T_{0}^{2} \right] +\mathrm{h}.\mathrm{c}..
\end{align}
Finally, the corresponding expectation value of the above Bell operator can obtained by Tr$[\rho \mc{B}]$ using the form of the density matrix in Eq.~\ref{eq:rho}. This gives us what we defined as our observable quantity \(\mc{I}_3\):
\begin{equation}
    \mc{I}_3 = \frac{1}{36}\left( 18+16\sqrt{3}-\sqrt{2}\left( 9-8\sqrt{3} \right) A_{2,0}^{1}-8\left( 3+2\sqrt{3} \right) C_{2,1,2,-1}+6C_{2,2,2,-2} \right) 
    \label{eq:i3}
\end{equation}
This observable is used to test whether the CGLMP inequality is violated or not in pair of qutrit system. Based on classical deterministic theory, the above operator is bounded up to 2. Any value that exceeds 2 indicates the violation of the Bell inequality given in Eq.~\ref{eq:i3}.

\section{Numerical Simulation}
\label{sec:result}
The signal process we choose are the semi-leptonic decay \(e^+e^- \rightarrow ZH, Z\rightarrow jj\) and leptonic \(e^+e^- \rightarrow ZH, Z\rightarrow\nu\bar{\nu}\) processes like Feynman diagrams shown in Fig.~\ref{fig:signal}. In our analysis, both the signal and background events are generated with \MGMCatNLO~\cite{Frederix:2012ps,Alwall:2014hca} at the parton-level, then showered and hadronized through \textsc{Pythia} 8.3 ~\cite{Bierlich:2022pfr}. The model is the default standard model, \delphes~\cite{deFavereau:2013fsa} version 3.0 is used to simulate detector effects with the settings for the CEPC detector~\cite{cepccard}. Jets are clustered from the reconstructed stable particles (except electrons and muons) using \textsc{FastJet}~\cite{Cacciari:2011ma} with the $k_{T}$ algorithm with a fixed cone size of $R_{jet} = 0.5$. Here we focus on the leptonic and semi-leptonic signal process. As for the background, we consider the following three processes as our corresponding background events:
\begin{itemize}
\item $e^{+}e^{-} \rightarrow \PZ\PZ$
\item $e^{+}e^{-} \rightarrow \PZ\PZ\PZ$
\item $e^{+}e^{-} \rightarrow \ell^{+}\ell^{-}\PH$
\end{itemize}
We selected these main backgrounds with the similar final state topology as the signal process, we also consider the cross section and the final states should include four leptons with two jets or with $\met$ after decaying. The leading order Higgs production through Higgsstrahlung can provide the largest cross section at $250~\PGeV$ collision energy. However, the final cross section for the signal process is suppressed due to the small branching ratio of the Higgs boson decay to the four leptons (muon pairs or electron pairs). The final leptons coming from the $\PZ$ boson pairs can be identified as four muons, four electrons or two electrons and two muons. We use the lepton pairs and require the opposite charge in each lepton pairs to reconstruct $\PZ$ bosons. The largest invariant mass of final lepton pairs, close to the the real $\PZ$ mass, is identified coming from on-shell $\PZ$ boson. If that invariant is much smaller than the actual mass of the $\PZ$ boson, then the lepton pairs is identified as coming from off-shell $\PZ$ boson. This is how we differentiate the two Z bosons from each other. 

\begin{figure}
    \centering
    (a)
    \includegraphics[width=0.4\linewidth]{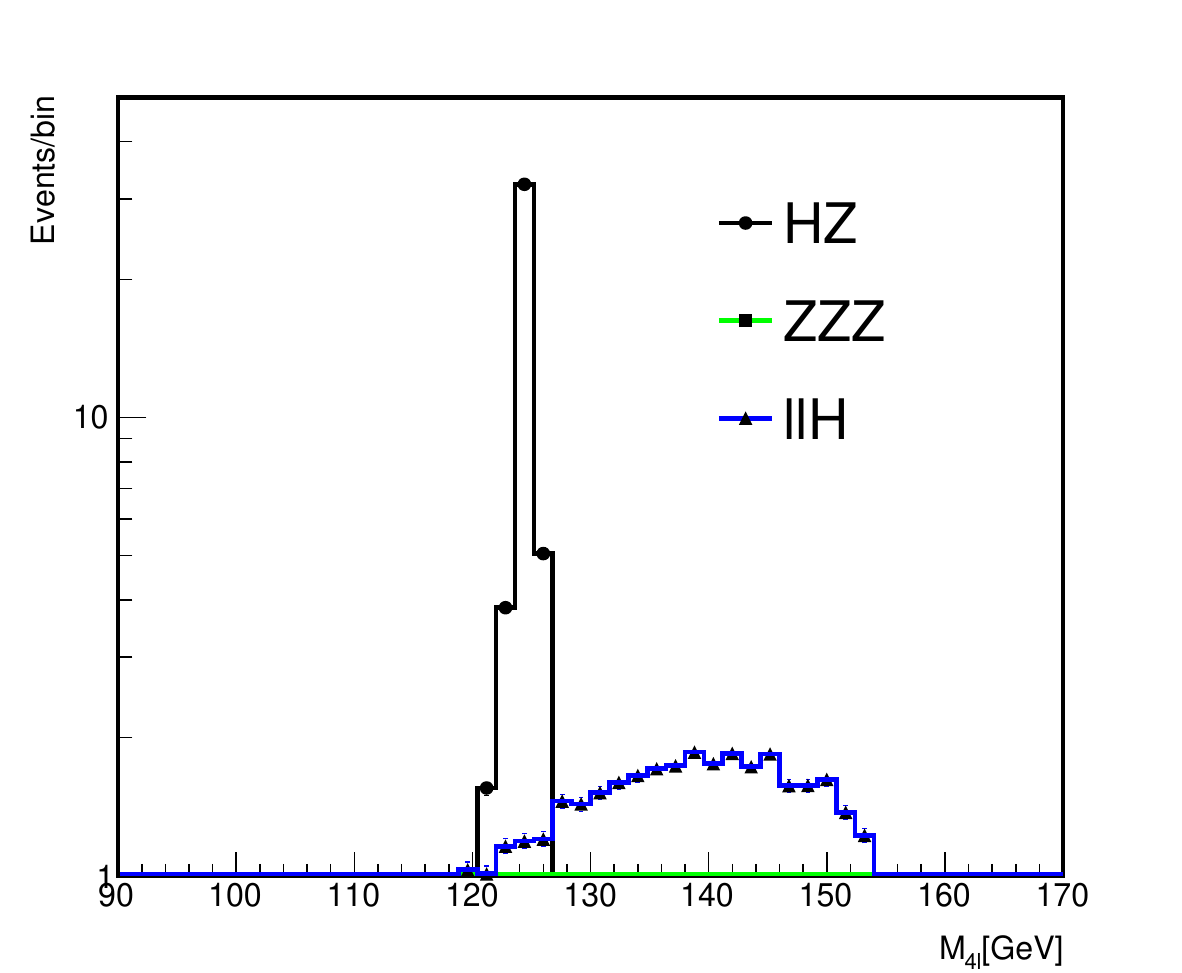} 
    (b)
    \includegraphics[width=0.4\linewidth]{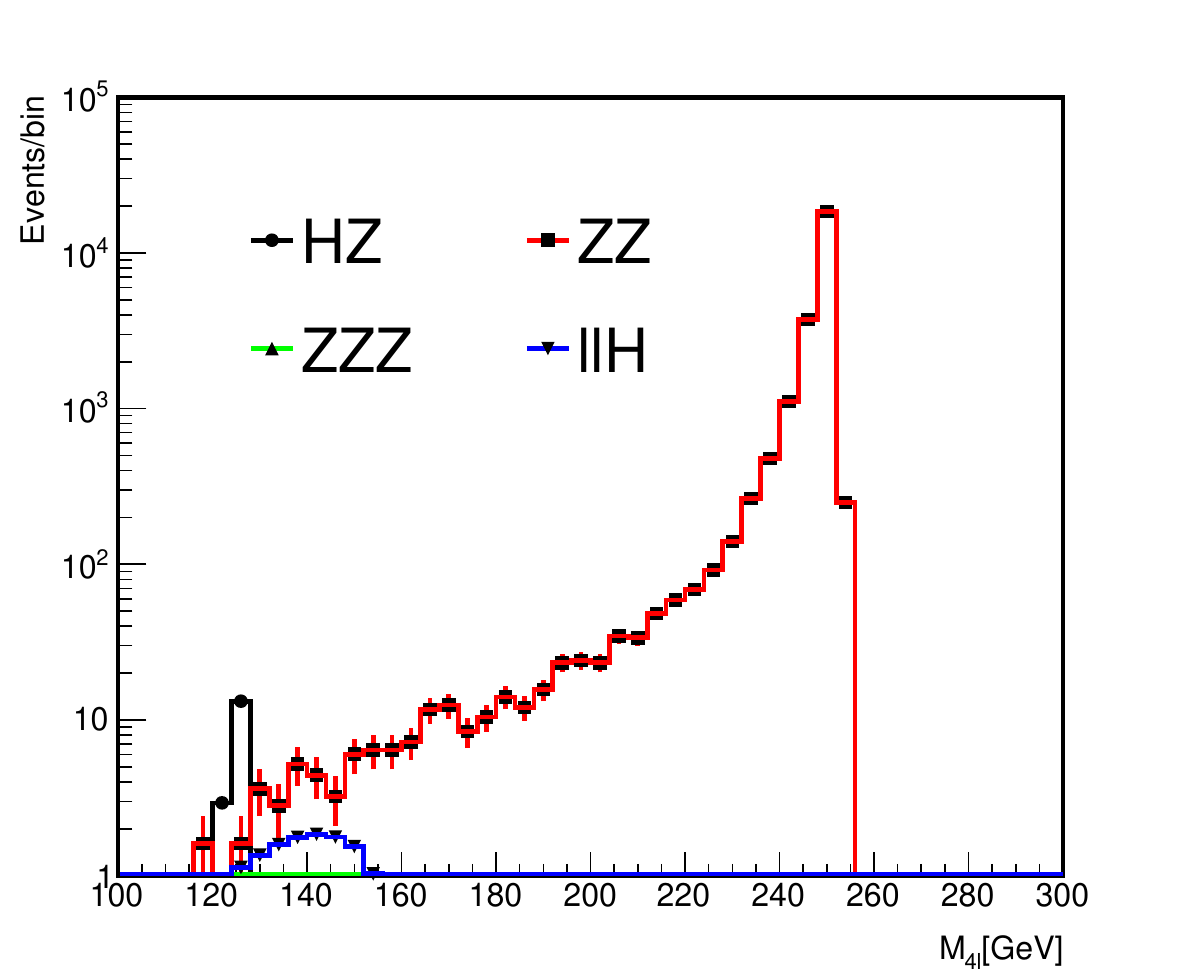}
    \caption{The distribution contains signal and background events as four leptons invariant mass. (a)semi-leptonic process (b) leptonic process. These results were obtained at a luminosity of 50 \(ab^{-1}\)}
    \label{fig:semi_m41}
\end{figure}

We consider the integrated Luminosity is $50~\rm{ab}^{-1}$, and the collision energy is 250 GeV, and we obtain the typical variable $\mathrm{M}_{4\ell}$ distribution for pure-leptonic channel and semi-leptonic channel shown in Fig.~\ref{fig:semi_m41}. As shown, the background is suppressed significantly relative to the signal, so that we can ignore it during the analysis of $\mc{I}_3$ and the coefficients $C_{2,1,2,-1}$ and $C_{2,2,2,-2}$.

In our analysis, we set a series of pseudo-experiments according to the expected number of events corresponding to target luminosity. The statistical uncertainties of the coefficients $\mc{I}_3, C_{2,1,2,-1}$ and $C_{2,2,2,-2}$ are dependent on the number of these pseudo-experiments, and we can ignore the systematic uncertainty because of the very clean final states in the lepton collider. The central values are calculated using Eq.~\ref{eq:ACLM}. The mean and the standard deviation ban be obtained through repeating the procedure over these pseudo-experiments. The observed value of the correlation coefficients and $\mc{I}_3$ change with respect to  the lower limit of the off-shell $\PZ$ boson.

The final measurements of the observable quantities of the pure-leptonic and semi-leptonic channels are shown in Table.~\ref{tab:result_lep} and Table.~\ref{tab:result_semi} with four different lower mass limits $M^{*}_{\PZ} \in [0, 10, 20, 30]\PGeV$. The mean value of $\mc{I}_3$ becomes larger with higher $M_{\PZ}^{*}$ as  expected and this means the $ZZ$ states  entangled more. However, the statistical uncertainties also rise as the $M^{*}_{\PZ}$ mass gets larger because of less events per pesudo-experiment. The non-zero value of the correlation coefficients, $C_{2,1,2,-1}$ and $ C_{2,2,2,-2}$, indicate that the two $\PZ$ boson states are entangled and this can be probed up to $2\sigma$ of significance in semi-leptonic channel and  $1\sigma$ of significance in pure-leptonic channel with lower $M^{*}_{\PZ}$ cut. The significance of the Bell inequality violation can reach more than $1\sigma$ in the semi-leptonic channel while below 1 $\sigma$ in the leptonic channel. 

\begin{table}[htb]
    \centering
    \caption{The Numerical result of the observable \(\mc{I}_3\) for the signal processes. (Semi-leptonic) \(\mathcal{L}=50 \mathrm{ab}^{-1}\)}
    \label{tab:result_semi}
    \begin{tabularx}{\textwidth}{l|X|X|X}
        \toprule
        \(M_z^*\) [\PGeV]  &  \(\mc{I}_3\) & \(C_{212-1}\) & \(C_{222-2}\) \\
        \midrule
        0 & \(2.823 \pm 0.640 (1.29\sigma)\) & \(-1.080 \pm 0.420 (2.57\sigma)\) & \(0.637 \pm 0.559 (1.14\sigma)\)  \\
        10 & \(2.913 \pm 0.692 (1.32\sigma)\) & \(-1.126 \pm 0.451 (2.50\sigma) \) & \(0.677 \pm 0.598 (1.13\sigma)\)  \\
        20 & \(3.092 \pm 0.800 (1.37\sigma)\) & \(-1.225 \pm 0.514 (2.38\sigma)\) & \(0.761 \pm 0.734 (1.04\sigma)\)  \\
        30 & \(3.048 \pm 1.816 (0.58\sigma)\) & \(-1.160 \pm 1.192 (0.97\sigma) \) & \(0.875 \pm 1.338 (0.65\sigma)\)  \\
        \bottomrule
    \end{tabularx}
\end{table}

\begin{table}[htb]
    \centering
    \caption{The Numerical result of the observable \(\mc{I}_3\) for the signal processes. (Leptonic) \(\mathcal{L}=50 \mathrm{ab}^{-1}\)}
    \label{tab:result_lep}
    \begin{tabularx}{\textwidth}{l|X|X|X}
        \toprule
        \(M_z^*\)[\PGeV]   &  \(\mc{I}_3\) & \(C_{212-1}\) & \(C_{222-2}\) \\
        \midrule
        0 & \(2.713\pm 1.167 (0.61\sigma) \) & \(-1.008 \pm 0.745 (1.35\sigma)\) & \(0.608 \pm 0.931 (0.65\sigma)\) \\
        10 & \(2.780 \pm 1.328 (0.59\sigma)\)  & \(-1.044 \pm 0.849 (1.23\sigma)\) & \(0.644 \pm 1.038 (0.62\sigma)\) \\
        20 & \(2.936 \pm 1.455 (0.64\sigma)\)   & \(-1.119 \pm 0.940 (1.19\sigma)\) & \(0.754 \pm 1.083 (0.70\sigma)\) \\
        30 & \(3.016 \pm 2.465 (0.41\sigma)\) & \(-1.129 \pm 1.616 (0.70\sigma)\) & \(0.905 \pm 1.617 (0.56\sigma)\) \\
        \bottomrule
    \end{tabularx} 
\end{table}

\section{Summary}
\label{sec:summary}
In this paper, we investigate the potential of probing quantum entanglement and the violation of the Bell inequality (CGLMP) in the Higgsstrahlung process $\PH \rightarrow \PZ\PZ \rightarrow 4\ell$ at a CEPC. All simulations are done with $\sqrt{s} = 250 \PGeV$ and $\mathcal{L} = 50~ab^{-1}$. Both on-shell and off-shell $\PZ$ bosons are reconstructed by the invariant mass of the lepton pairs. We focus on two signal process: semi-leptonic and pure-leptonic channel according to different final states. These signals are so clean that the background to them can be safely ignored.

Because of the spin-zero property of the Higgs boson, the $\PZ\PZ$ system arising from Higgs decay is in a spin-singlet state, which is maximally entangled. This can reduce the number of free parameters in the polarization density matrix of the joint $\PZ\PZ$ system, giving only two independent parameters: $C_{2,1,2,-1}$ and $C_{2,2,2,-2}$. The density matrix of that system is parametrized using irreducible tensors~\cite{Aguilar-Saavedra:2022wam}. Measuring the spin-correlation coefficients $C_{2,1,2,-1}$ and $C_{2,2,2,-2}$ by  determining the spherical coordinates of the four leptons in the final states enables us to obtain the density matrix of the joint system and probe the presence of quantum entanglement. Any non-zero value of either $C_{2,1,2,-1}$ or $C_{2,2,2,-2}$ can prove $\PZ\PZ$ state is an entangled quantum state. Quantum entanglement can be measured with a significance up to $2\sigma$ in semi-leptonic signal channel and 1 $\sigma$ in pure-leptonic signal channel. In the end, the significance of the Bell inequality can be probed up to $1\sigma$ in semi-leptonic channel.


\begin{acknowledgments}
This work is supported in part by the National Natural Science Foundation of China under Grants No. 12150005, No. 12075004, and No. 12061141002, by MOST under grant No. 2018YFA0403900.
\end{acknowledgments}

\bibliographystyle{ieeetr}
\bibliography{refs}
\end{document}